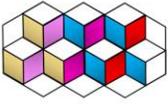 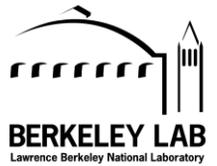

# Quantifying and Visualizing Hidden Preferential Aggregations Amid Heterogeneity

David H. Nguyen, PhD
Principal Investigator
Tissue Spatial Geometrics Lab

Affiliate Scientist
Dept. of Cellular & Tissue Imaging
Division of Molecular Biophysics and Integrated Bioimaging
Lawrence Berkeley National Laboratory
DHNguyen@lbl.gov

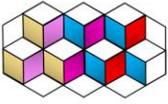

## Abstract

Biological systems often exhibit a heterogeneous arrangement of objects, such as assorted nuclear chromatin patterns in a tumor, assorted species of bacteria in biofilms, or assorted aggregates of subcellular particles. Principle Component Analysis (PCA) and Multiple Component Analysis (MCA) provide information about which features in multidimensional data aggregate, but do not provide *in situ* spatial information about these aggregations. This paper outlines the Numericized Histogram Score (NHS) algorithm, which converts the histogram distribution of shortest distances between objects into a continuous variable that can be represented as a spatial heatmap. A histogram can be transformed into an intensity value by assigning a weighting factor to each sequential bin. Each object in an image can be replaced by it's NHS value, which when calibrated to a color scale results in a heatmap. These spatial heatmaps reveal regions of aggregation amid heterogeneity that would otherwise mask loco-regional spatial associations, which will be especially useful in the field of digital pathology. In addition to visualizing aggregations as heatmaps, the ability to calculate degrees of recurring patterns of aggregation allows investigators to stratify samples for further insights into clinical outcome, response to treatment, or "omic" subtypes (genomic, transcriptomic, proteomic, metabolomic, etc.).

## Graphical Abstract

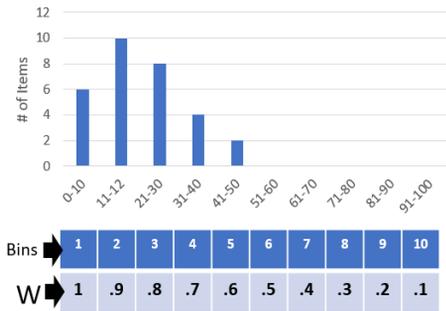

The Numericized Histogram Score (NHS) Algorithm

$$S = \frac{\sum_{i=1}^{n}[B_i \times W_i]}{M} = \frac{\sum_{i=1}^{n}[Q_i]}{M}$$

S = Saturation Score

B = Number of items in a bin

W = Proximity Weight assigned to each bin

Q = [B x W] = Adjusted Proximity Score

M = Maximum Possible Adjusted Proximity

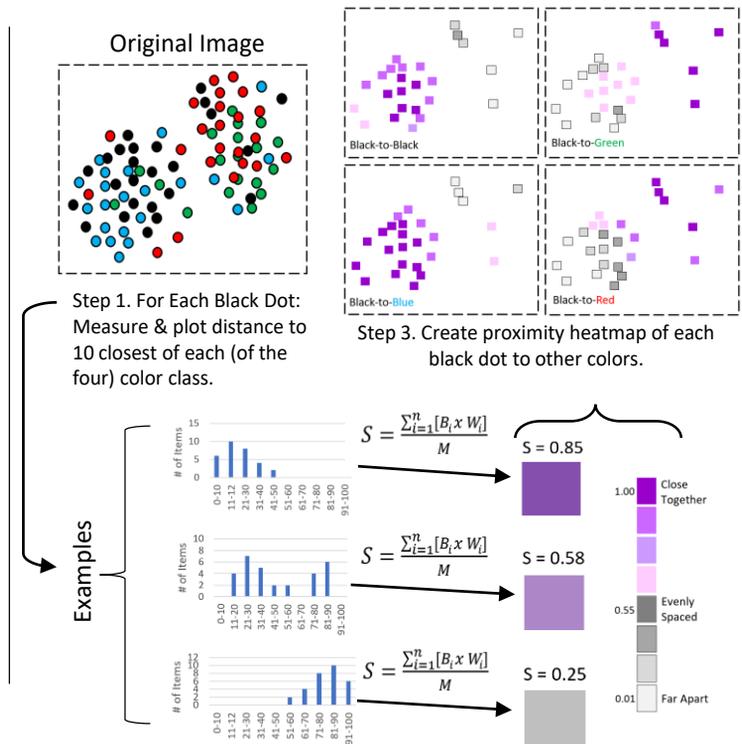

Step 1. For Each Black Dot: Measure & plot distance to 10 closest of each (of the four) color class.

Step 3. Create proximity heatmap of each black dot to other colors.

Step 2. Numericize each histogram to get a value representing proximity.

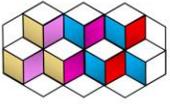

## Background Information

Biological systems often have a high degree of heterogeneity, such that it is difficult to decipher whether certain objects tend to aggregate with each other but apart from other objects. Examples of such systems include the aggregation species of bacteria in soil or in the intestine; different subpopulations of cells in a tumor; or the preferential interaction of certain species of insects, birds, or other organisms in an ecosystem.

Each species of bacteria or each subpopulation of cells in a tumor represents a different class of object. For example, if a tumor has six different types of cells that can be distinguished from each other, then there are six different classes of objects in an image of this tumor. In this paper, a system of four different object classes will be analyzed as proof of concept (Figure 1).

### Figure 1

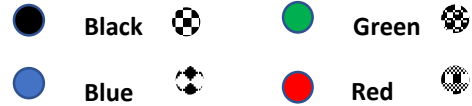

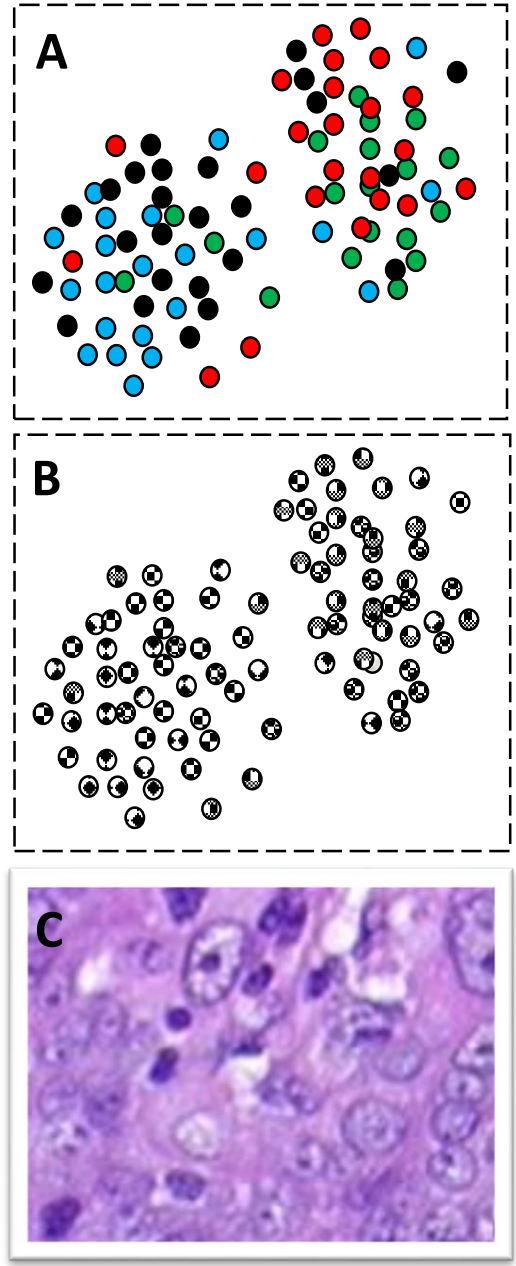

**Figure 1. Heterogeneity Masks Patterns of Aggregation.** Panel A and B are the same, except for the color and patterning of each class of objects (see legend). The complexity of finding spatial order within B is akin to finding order in C, which is a human breast tumor stained by hematoxylin & eosin (H&E).

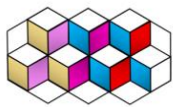

The ability to measure preferential aggregations allows investigators to find insights about how a biological system organizes itself. Since structure and function go hand-in-hand, the ability to stratify specimens by the degree of loco-regional orderliness within their heterogeneous patterns will improve the diagnosis of cancer and the understanding of why certain patients don't respond to certain treatments. The ability to measure preferential aggregations will also allow investigators to learn about the behaviors of one species in an ecosystem in relation to another species, which would remain hidden without a high degree of quantitation.

**Method**

Objects that tend to aggregate will statistically, on average, be located closer to each other than those that do not. The proof of concept image in this paper has four different classes of objects (Figure 1): black, blue, red, and green. For each object in each class, the shortest distance between that object and the nearest 10 objects of another class are measured. For example, the black class has 25 individual objects. Figure 2 shows the shortest distance between object #17 of the black class and its 10 nearest neighbors in the blue, red, green, or black classes. The investigator can choose to measure the distance of each object to any number of nearest neighbors, depending on the desired degree of statistical rigor and prior knowledge of the biological system. In this case, 10 neighbors was arbitrarily chosen. This approach is repeated for every object in the black class, resulting in four sets of distance measures for the image: black-to-black, black-to-blue, black-to-red, and black-to-green. A complete assessment of the image in Figure 2A would result in 10 sets of distance measures representing all possible combinations of color pairs, and each color to itself (i.e. Do black objects tend to aggregate with themselves?): black-to-black, black-to-blue, black-to-red, and black-to-green; blue-to-red and blue-to-green; red-to-green; blue-to-blue, red-to-red, green-to-green. All measurements in this paper were done using ImageJ/Fiji open-source software.

The results from measuring shortest distances of each object to its 10 nearest neighbors is then plotted as histograms that have the same bin sizes (Figure 3). The histograms reveal what is visually obvious, that object black #17 aggregates with other black objects and other blue objects, while being far away from red or green objects. However, in order to visualize such aggregation patterns in biological systems that do not have easily identifiable classes (Figure 1A, 1B), there needs to be an algorithm to numericize the distinct shapes of each histogram.

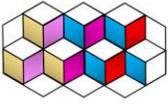

**Figure 2**

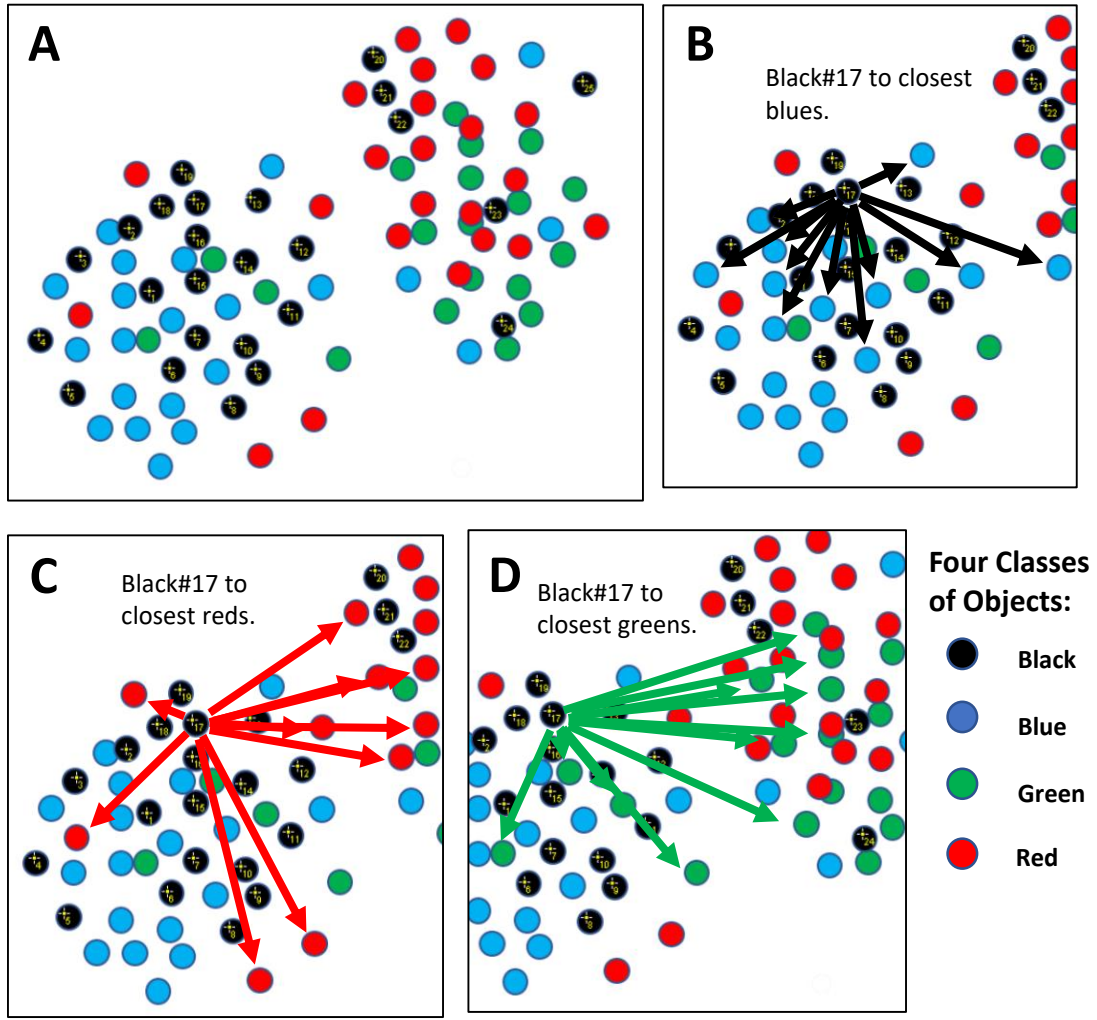

**Figure 2. Trends of aggregation can be quantified by systematically measuring shortest distance between objects.** (A) Image showing the distribution of four classes of objects (black, blue, red, and green). (B) Lines showing shortest distances between black dot #17 to the 10 closest black dots. (C) Lines showing shortest distances between black dot #17 to the 10 closest red dots. (D) Lines showing shortest distances between black dot #17 to the 10 closest green dots.

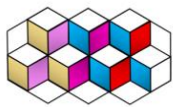

Each of the histograms in Figure 3 can be converted to a continuous variable ranging from 0.0 to 1.0 (Figure 5). This allows for the histogram representing each object in an image to be represented as a value that saturates at an upper limit of 1.0. In other words, each object in an image can be represented by a pixel whose intensity and/or color is within the scale of 0.0 to 1.0, which results in a heatmap of the original image (Figure 6).

By adding a sequential weighting factor, referred to as "W," to each bin in a histogram the equation in Figure 4B captures the "shape" of a distribution. In this case, the histograms had 10 bins with the first ones representing the shortest distances between two objects. This first bin was assigned a value of 1.0, while each subsequent bin had a W value of 0.1 less than the previous bin. The W scale in this case was designed to accentuate the significance of histograms that had many items in the bins with the shortest distances, since the equation in Figure 4B is meant to measure degrees of proximity. This equation will be referred to as the Numericized Histogram Score (NHS) algorithm. The numericized histogram (NH) score is represented by "S," which stands for saturation and ranges between 0.0 and 1.0. Saturation and transparency are inversely related to each other, so the higher the S score of a histogram the more dense (i.e. less transparent) it will appear as an area of color in a heatmap.

Each bin in a histogram has two factors: "B," the number of items in that bin, and W, the weight assigned to that bin. Multiplying B and W for a bin produces "Q," the adjusted proximity value for that bin. By summing all Q's of a histogram (one Q for each bin), the shape of the histogram is captured in one value. The NHS algorithm then introduces a normalization factor "M," which is the maximum possible "closeness" of the histogram. M is derived by assuming that all items in a histogram are in the first bin, which in this paper represents the shortest distances. Dividing Q by M results in the saturation score S (Figure 4B).

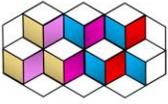

**Figure 3**

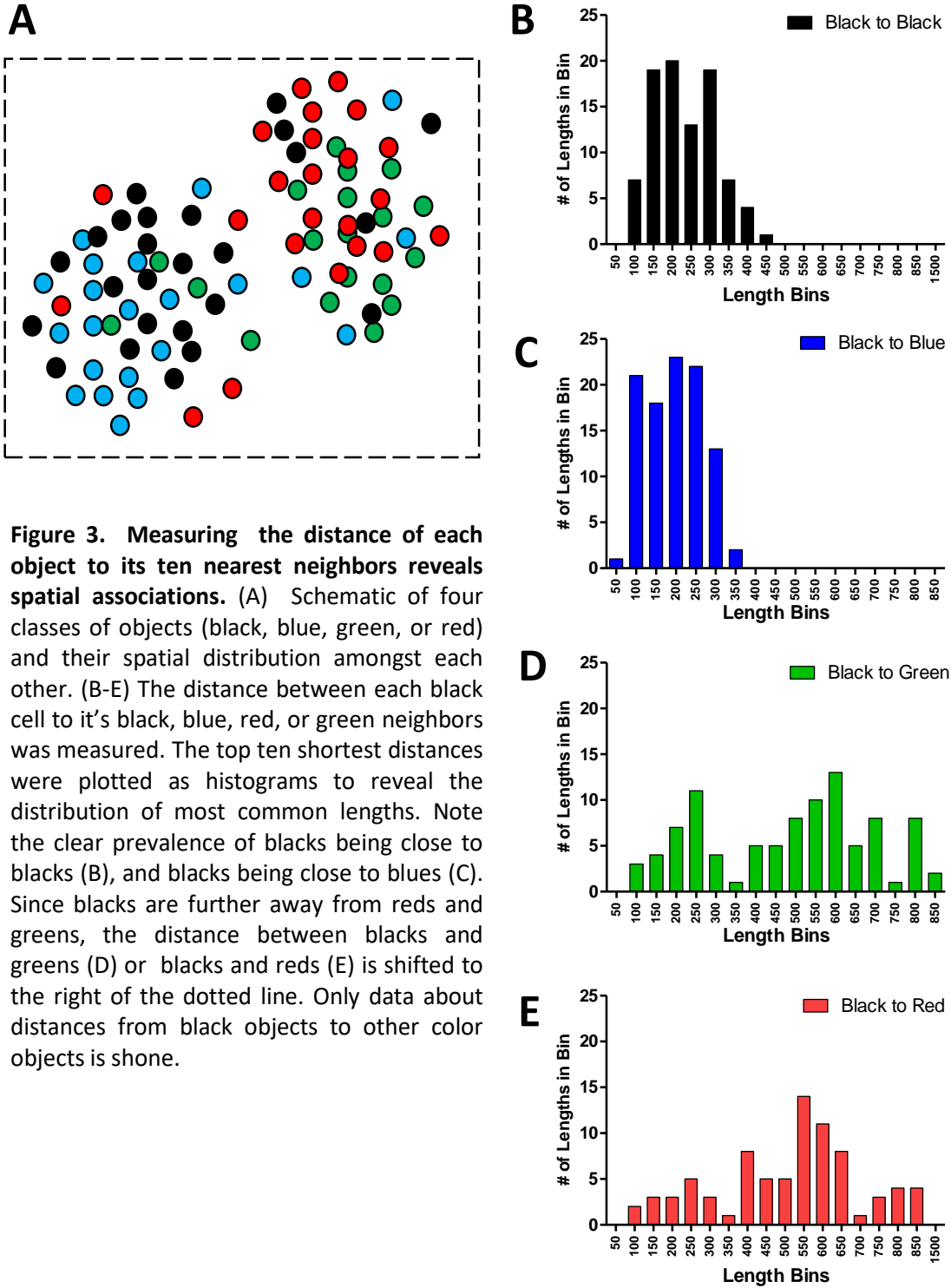

**Figure 3. Measuring the distance of each object to its ten nearest neighbors reveals spatial associations.** (A) Schematic of four classes of objects (black, blue, green, or red) and their spatial distribution amongst each other. (B-E) The distance between each black cell to it's black, blue, red, or green neighbors was measured. The top ten shortest distances were plotted as histograms to reveal the distribution of most common lengths. Note the clear prevalence of blacks being close to blacks (B), and blacks being close to blues (C). Since blacks are further away from reds and greens, the distance between blacks and greens (D) or blacks and reds (E) is shifted to the right of the dotted line. Only data about distances from black objects to other color objects is shone.

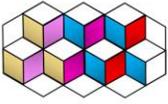

**Figure 4. The Numericized Histogram Score (NHS) Algorithm for Quantifying the Shape of a Binned Distribution.** (A) A histogram showing that the majority of items are in the smaller bins, which in the context of this paper are the bins containing the short distances. In this paper, this histogram represents the 10 nearest neighbors of one color class (i.e. blue) to one object in another color class (i.e. black #17 from Figure 2). (B) The rightward (tail extends towards the left) skew of the histogram in panel A can be quantified by assigning a weighting factor to each bin along the X-axis of the histogram. The weighting factor in this case was designed to give histograms with a rightward skew a larger value than those that skewed leftward. The NHS algorithm takes into consideration the number of items in a bin ("B"), the location of the bin on the x-axis, the weighting factor that defines the significance of the bin position ("W"), and the total possible degree of skewing if all items were in the first bin ("M"). "M" serves as a normalizing factor that keeps the range of "S" between 0.0 to 1.0.

Figure 4

A

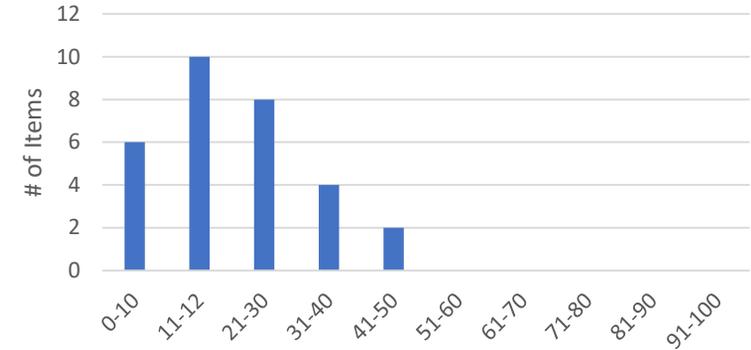

B

**The Numericized Histogram Score (NHS) Algorithm**

$$S = \frac{\sum_{i=1}^{n}[B_i \times W_i]}{M} = \frac{\sum_{i=1}^{n}[Q_i]}{M}$$

S = Saturation Score

B = Number of items in a bin

W = Proximity Weight assigned to each bin

Q = [B x W] = Adjusted Proximity Score

M = Maximum Possible Adjusted Proximity. [M = $W_1$ x (Total # of Items)]

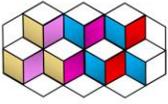

Figure 5 describes two options for visualizing NH scores as heatmaps. Since NH scores, represented as S, are within the range of 0.0 to 1.0 they can be represented as varying degrees of transparency of a single color. However, to make visualization more effective it is beneficial to create a threshold around which two colors can be based. This effort is attainable by calculating S for the hypothetical situation in which items in a histogram are evenly distributed across all bins, resulting in a (flat) uniform distribution ($S_{UD}$). The result is $S_{UD}$, which is a threshold for changing the color that represents distances that are "short" or "long."

Option 2 in Figure 5 describes a method to increase the range of difference between S scores. This can be done by transforming all S scores by the same function (for example, $y = 400x^2$), which expands the range of possible values for more contrast between shades of a single color. Subtracting the transformed S score of the hypothetical uniform distribution situation from every other S score results in a color scale that is centered around 0. This allows both "short" and "long" distances to be represented as increasingly intense or bright colors.

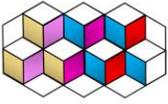

**Figure 5**

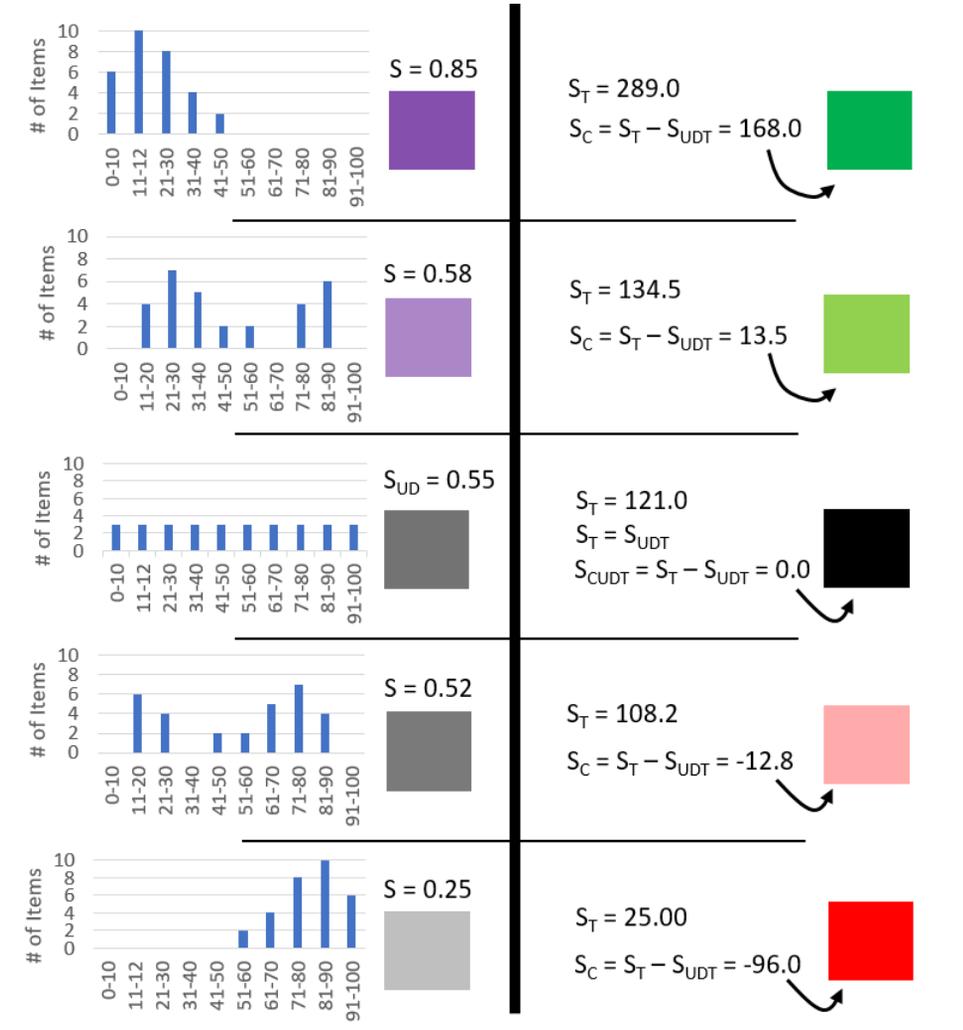

**Figure 5. Options for Visualizing the Numericized Histogram Score (NHS) Value.** S, the saturation score; $S_T$, the saturation score transformed by $y=400x^2$ or other method to accentuate range; $S_{UD}$, the saturation score of a uniform distribution, which is a hypothetical situation that is treated as a threshold value wherein all items are evenly distributed across all bins; $S_{UDT}$, the transformed saturation score of a uniform distribution; $S_C$, the centered saturation score resulting from the following subtraction, $S_T - S_{UDT}$; $S_{CUDT}$, the centered hypothetical threshold cut-off value assuming that items are evenly distributed across all bins (results from $S_T - S_{UDT}$ and is always 0).

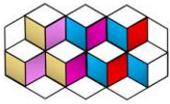

**Figure 6**

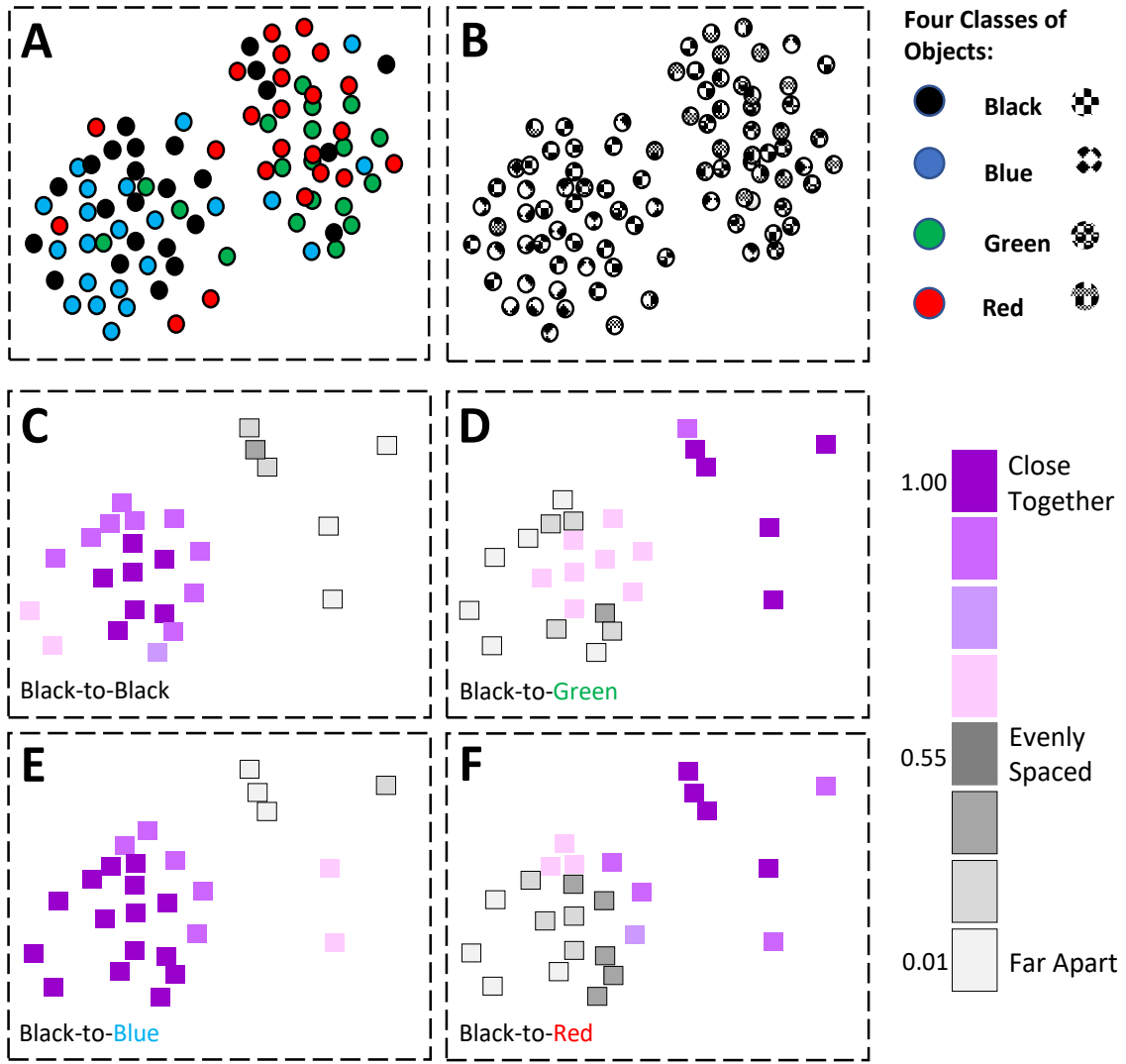

**Figure 6. Hypothetical Case of a Proximity Heatmap Based on the Numericized Histogram Score (NHS) Algorithm.** (A & B) These two panels are the identical arrangements of four classes of objects. The colors in panel A make the aggregations between black & blue, and red & green obvious to the naked eye. However, the textured patterns in panel B masks the aggregation patterns. The complexity in panel B is similar to what a pathologist sees when viewing tissues under a microscope. (C-F) Proximity heatmaps produced based on the saturation score (S) that resulted from the NHS algorithm. The density and intensity of squares reveals which regions contain aggregations (purple shades) or repulsions (grey shades) between black objects and green, blue, red, or other black objects.

## Potential Applications of Numericized Histogram Score (NHS) Algorithm

The initial purpose for generating the NHS algorithm was to develop a method for pathologists to visualize aggregations of nuclear subtypes in sections of tumors stained by H&E (Figure 7). However, the NHS algorithm will be useful to any field of science that analyzes images that have different classes of objects arranged in a non-uniform pattern (Figure 8).

**Figure 7**

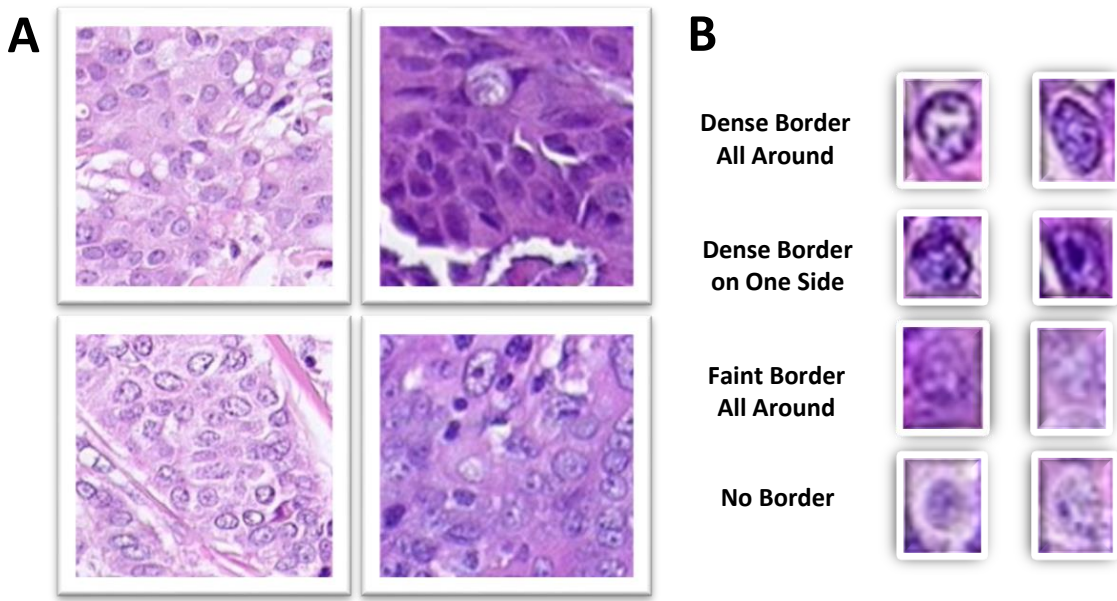

**Figure 7. Hematoxylin & Eosin staining of human breast tumors.** (A) Condensed chromatin appears dark purple/blue when stained by hematoxylin and eosin. (B) Distinct chromatin patterns that co-exist in human breast tumors. Pathologists determine the grade of a tumor by factors that include the nuclear morphology and the density of chromatin staining. Without computational approaches, however, pathologists cannot give refined diagnoses based on minute but recurring chromatin patterns.

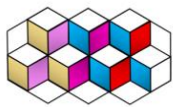

**Figure 8**

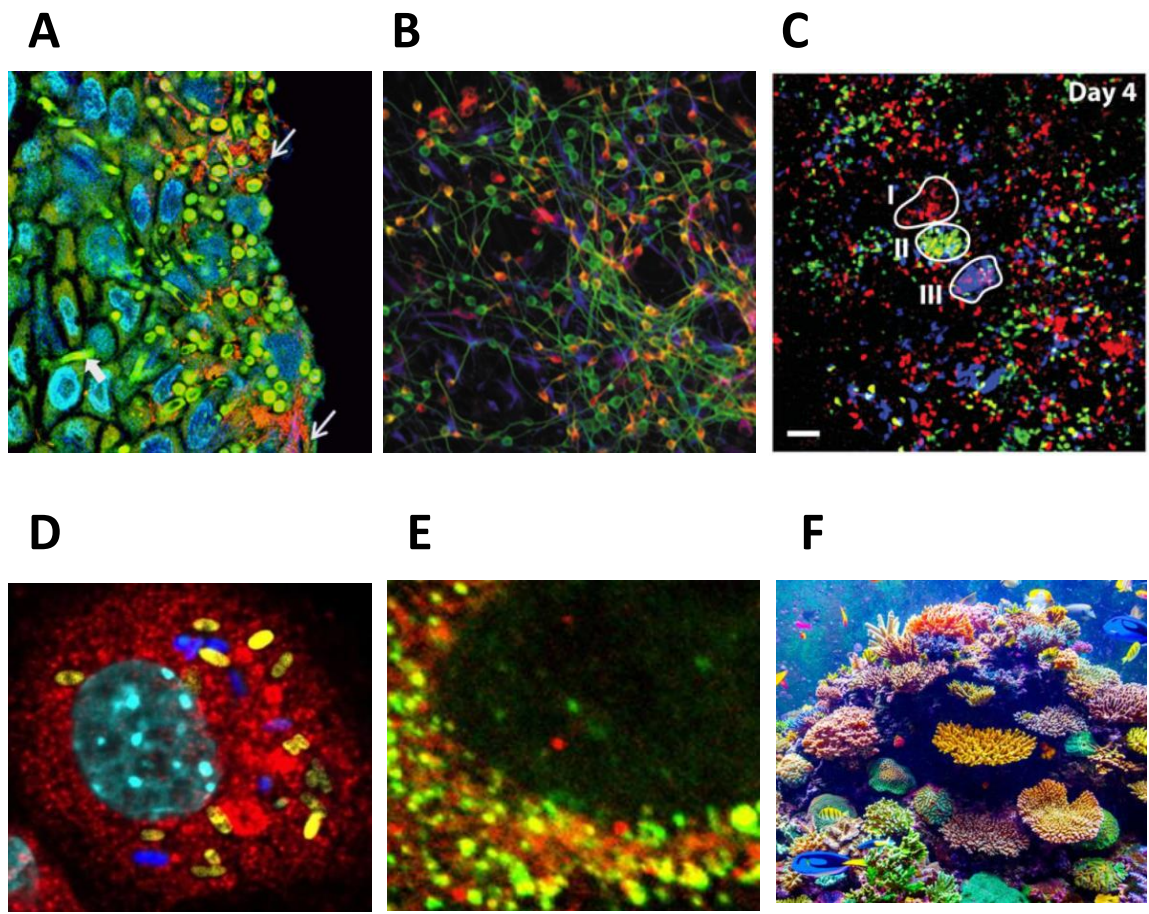

**Figure 8. Other types of image data that can benefit from the NHS algorithm.** (A) Microbial invasion of epithelial linings: bacteria, red; yeast, green (from Cavalcanti, 2015). (B) Neuronal progenitor interactions (from Pflumm, 2017). (C) Cellular lineage tracking in tumors: red, green, and blue are cancer cells that originate from distinct cells of origin (from Zomer, 2013). (D) Immune cell interactions with pathogens: *E. coli* are yellow and blue inside of a red macrophage (from Public Library of Science, 2013). (E) Subcellular localization of proteins: fluorescently labeled GLUT8 transporter protein in the cytosol (red and green) surrounding the nucleus (from Schmidt, 2006). (F) Distribution of coral reef fauna and flora (from Porterfield, 2016).

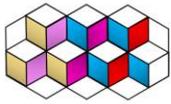